\documentclass[a4paper,11pt]{article}
\usepackage{amsmath}
\usepackage{amssymb}
\usepackage{amsthm}
\usepackage{thmtools}
\usepackage[round]{natbib}
\usepackage{color}
\usepackage{graphicx}
\usepackage{hyperref}
\hypersetup{
    unicode=false,          
    pdftoolbar=true,        
    pdfmenubar=true,        
    pdffitwindow=false,     
    pdfstartview={FitH},    
    colorlinks=true,       
    linkcolor=black,          
    citecolor=blue,        
    filecolor=magenta,      
    urlcolor=blue           
}
\usepackage{algorithmic}
\usepackage{algorithm}
\usepackage{adjustbox}
\usepackage{textcomp}
\usepackage{tikz}
\usetikzlibrary{calc}
\usepackage{fullpage}
\usepackage{authblk}
\usepackage[output-decimal-marker={\cdot}]{siunitx}

\title{Modelling and predicting the spatio-temporal spread of Coronavirus disease 2019 (COVID-19) in Italy}

\author[1]{Diego Giuliani}
\author[1]{Maria Michela Dickson}
\author[1]{Giuseppe Espa}
\author[2]{Flavio Santi}

\affil[1]{Department of Economics and Management, University of Trento}
\affil[2]{Department of Economics, University of Verona}

\begin{document}
\maketitle

\section{Introduction}

Since the first cases occurred in Wuhan (China), Coronavirus disease 2019 (COVID-19) caused by SARS-CoV-2 virus, has raised serious concern. The virus, which made the inter-species jump to humans probably from bats through another intermediate animal host~\citep{li2020c}, causes a severe respiratory syndrome characterized by a strong person-to-person contagiousness via aerial route~\citep{guan2020,li2020a}. Not surprisingly, SARS-CoV-2 has spread throughout and outside of China in a short time following an uneven spreading pattern, despite the severe control measures put in place in the Wuhan region. Italy is the second most affected country in the World and the first in Europe, with \num{35713} confirmed cases according to the Italian Department of Civil Protection as of 18~March~2020. The reasons for Italy's strong involvement are not yet clearly identified, especially because it seems this was not the first European country involved in the epidemic~\citep{rothe2020}. Furthermore, the spread of COVID-19 in Italy has not followed a uniform pattern on the territory. After the first case not directly connected with China has been discovered on 20~February~2020 in the province of Lodi (north-west Italy), it has been possible to observe the diffusion of the disease in the three northern regions of Lombardy, Veneto and Emilia-Romagna.  It took some days to record contagions in other Italian regions, whether near or far, and even today that COVID-19 is spread throughout Italy, some provinces are more affected than others. Currently, \num{17713} out of \num{35713} confirmed cases (\num{49.6}\%) came from Lombardy region, while Lombardy, Veneto and Emilia-Romagna accounted for \num{25452} cases (\num{71.3}\%), as a whole. It took some days to record contagions in other Italian regions, whether near or far, and even today that COVID-19 is spread throughout Italy, some provinces are more affected than others.

Italian regions are extremely different to each other, especially between southern and northern regions, with the latter representing the industrial, economic and financial epicentre of the country. Furthermore, on the northern Italian territory insists large flows of people coming from other areas because of labour commuting, and also of people looking for a more efficient healthcare, which is located right in the area where the virus is occurring with greater force. These differences might have obvious in influences on the spread pattern of COVID-19. Strict measures to prevent and contain the epidemic, including social distancing, closure of businesses and schools, prohibitions of travel and going outdoor, were enforced starting from the cited regions and later applied to the entire country. It is therefore of striking importance to trying to understand the contagion phenomenon and to predict its spatial diffusion as well as its temporal trend.

Many scholars are trying to give a contribution to the problem, both in open online venues and on scientific publications, debating about reproductive number, mortality distributed for age and gender and, more in general, epidemic features, without neglecting to avail of mathematical and statistical models. The main contributions use deterministic epidemiological models~\citep[see][among others]{kucharski2020,liu2020,sun2020}, all focussing on the time evolution of the phenomena, especially with predictive purposes. The prediction of new infected, deceased and healed is certainly essential for health policy in order to estimate the capacity of a health system to cope with the stress caused by a pandemic. Nevertheless, in the recent literature the relevance of space in the diffusion of the COVID-19 is treated only marginally~\citep{chinazzi2020,li2020b}, although the importance of spatial and spatio-temporal autocorrelation in epidemiology was already highlighted in the seminal book by \cite{cliff1981}. The authors show the effectiveness of statistical methods in analyzing the incidence of some epidemics, such as e.g. measles in Cornwall, 1969--1970, cholera in London in 1849 and tuberculosis and bronchitis in Wales, 1959--1963. In the last 20 years spatial epidemiology has been evolving rapidly and it has found a great response in medical applied research~\citep[for a recent review on methods and applications in the field, see][]{kirby2017}.

The importance of the space in the study of disease transmission, like all natural phenomena, answers to the first law of geography~\citep{tobler1970}, for which ‘‘everything is related to everything else, but near things are more related than distant things''. Reformulating this concept, it could be said that the phenomenon external to an area of interest affects what goes on inside. In light of this, we found imperative to include a component that can account for spatial dependence among areal units in the study of COVID-19 diffusion, in order to explain the strength and direction of the spreading on the territory as well as in time. This is particularly relevant in countries such as Italy or Germany in which the local health systems interested by the disease are strongly regionalized. The territorial specificity, which can result in more or less drastic containment measures, cannot be neglected in a model construction that has the ambition to explain how the contagion moves in space and time and to allow reliable spatio-temporal predictions.

The purpose of the present paper is to model and predict the number of COVID-19 infections, drawing out the effects of its spatial diffusion. Forecasts about where and when the disease will occur may be of great usefulness for public decision-makers, as they give them the time to intervene on the local public health systems.

\section{Methods}

\subsection{Data}

The Italian Department of Civil Protection according with the Italian Ministry of Health, has started to release a daily bulletin about COVID-19 infections in Italy since 26~February~2020. By 3~March~2020 data were also released at NUTS-3 level (provinces),\footnote{NUTS (Nomenclature of Territorial Units for Statistics) is a European Union geocode standard for referencing the subdivisions of countries for statistical purposes, established by Eurostat in agreement with each member state. The level of NUTS are four, nested into each other: NUTS-0 are the sovereign countries; NUTS-1 are the major socio-economic regions; NUTS-2 are basic regions for the application of regional policies (Italian regions); NUTS-3 are small regions for specific diagnoses (Italian provinces).} having been previously released only at the highly aggregated NUTS-2 level. Data are available from the websites of the main Italian newspapers, such as, \textit{Il Sole 24 ore}.\footnote{ \url{https://lab24.ilsole24ore.com/coronavirus}.} Therefore, immediately after the release, we were able to collect the infectious disease count time series at provincial level for the period from 26~February~2020 to 18~March~2020 using two sources. Since spatial dependence may result stronger than it actually is at a coarse spatial resolution~\citep{arbia1996}, we have wanted to use data at the provincial level of aggregation, which is the freely available finest possible level of aggregation. For the first six days, data were provided only to media outlets and published by newspapers in the form of interactive maps. For this period, we have used web scraping techniques to track raw data upon which maps have been constructed. For the remaining days, we simply downloaded data about provinces released by the Department of Civil Protection. Clearly, at this stage, data suffer from several problems of quality. The number of contagions is continuously updated and positive swabs need to be confirmed by the National Institute of Health, not to mention several cases of delayed reporting by the local authorities. Therefore, the results presented in the paper may benefit from availability of more accurate data.

\begin{figure}[p]
\begin{center}
\caption{Daily counts of COVID-19 infections in Italy.}
\includegraphics[scale=0.40]{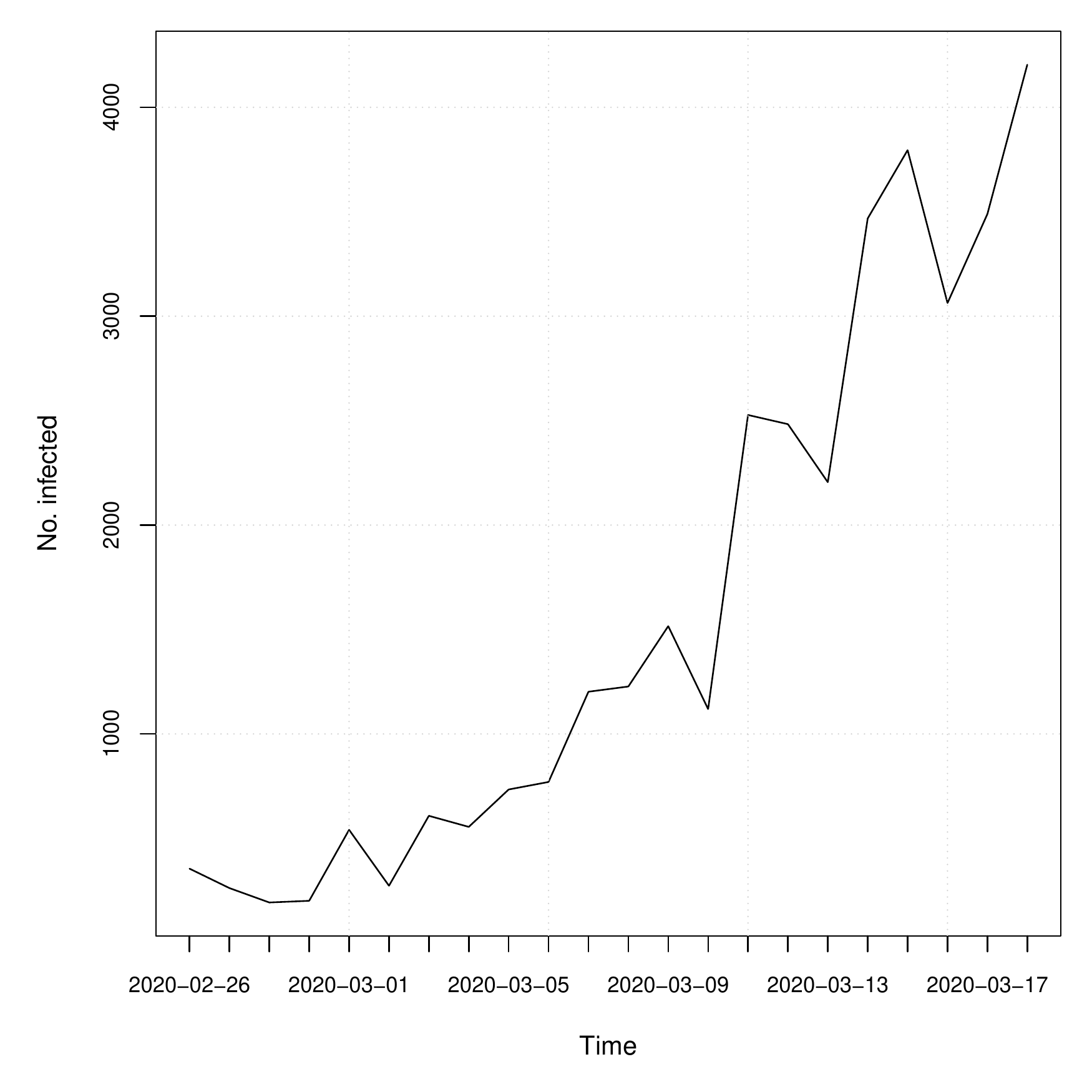}
\label{GlobalTS}
\end{center}
\end{figure}

\begin{figure}[p]
\begin{center}
\caption{Italian provinces coloured according to COVID-19 daily incidence (total number of infections per 1000 inhabitants) in the Italian provinces (26 February 2020 -- 18 March 2020).}
\includegraphics[scale=0.50]{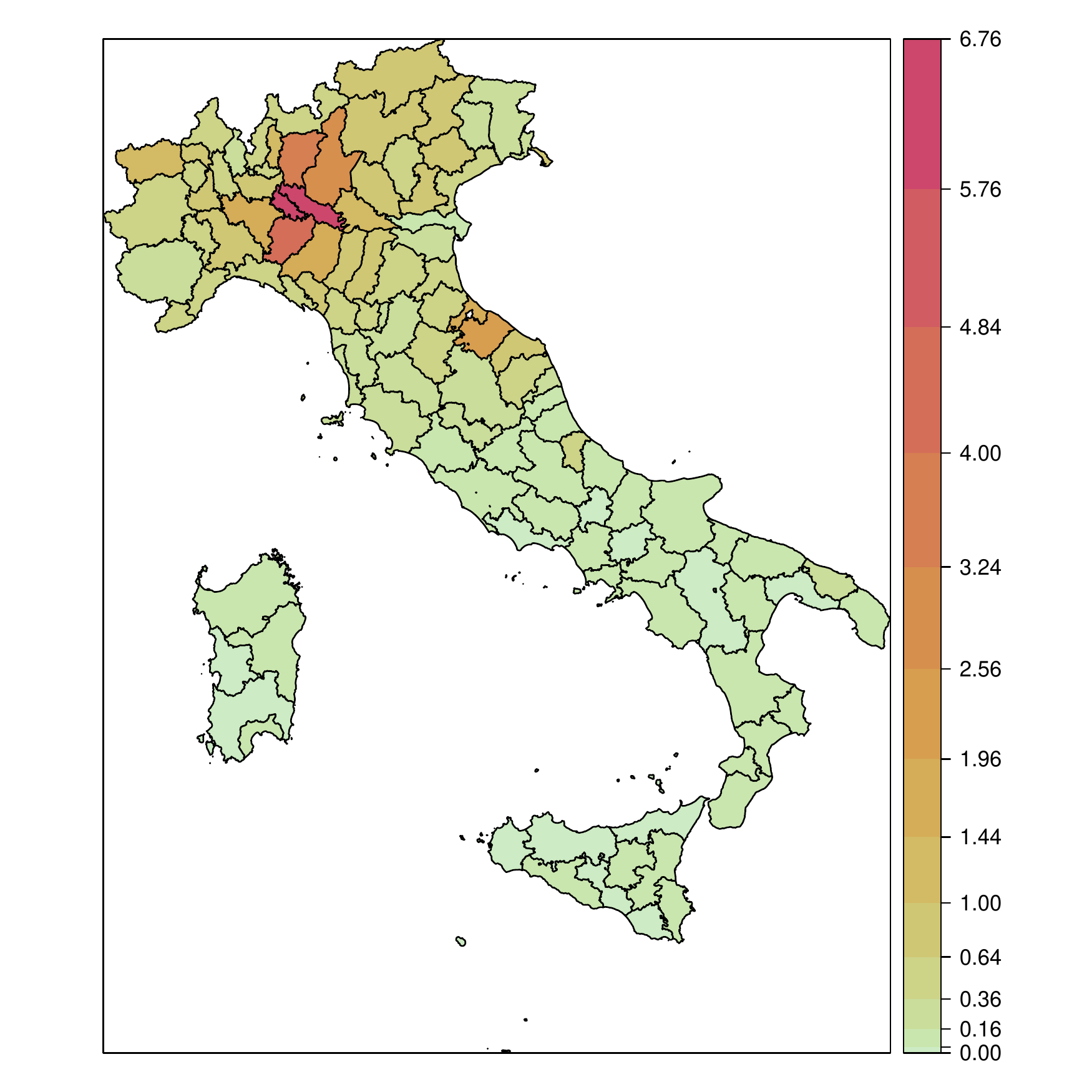}
\label{IncidMap}
\end{center}
\end{figure}

The overall temporal distribution of daily counts of COVID-19 infections is depicted in Figure~\ref{GlobalTS},\footnote{The overall time series here depicted is given by the aggregation of regional time series and may not correspond exactly to the national time series because some cases could not being assigned precisely to the relative province} while the spatial distribution is showed in the map in Figure~\ref{IncidMap}. The two plots indicate that while the growth of cases over time in Italy has been globally exponential, it is characterized by strong local heterogeneity and also by a relevant spatial pattern in the form of a hot-spot in the northern part of the country (the plot of the time series disaggregated for all Italian provinces is reported in the Appendix).

\subsection{Statistical analysis}

The evolution of the number of daily infections is studied by means of a statistical model which has been adapted from that proposed in \citet{HeldEtAl2005} and \citet{PaulHeld2011}, where both temporal dynamics of the infections and its geographical mechanism of spread are considered and estimated~\cite[see][for an implementation of the model in epidemiology]{AdegboyeEtAl2017,Cheng2016}.

The expected number of infections occurred in a province in the course of a day may be decomposed according to three additive components representing three different channels through which the contagions may increase.  

The first component of contagion is represented by the spread mechanism of the COVID-19 within each province, which depends on the cumulative incidence of the disease and determines the speed of diffusion of the virus in the future. This component determines the temporal dynamics of the contagion within each province and for this reason is referred to as epidemic-within component (see Appendix).

Another channel is attributable to contagions originated from other provinces, and in particular from provinces which are geographically close to each other. The origin of such spatial dimension of contagion can be found in the high mobility of people across provinces, and substantially contributes to explain the observed diffusion of the virus from a limited number of focuses to a wider area which, at the time we are writing, heavily affects the most part of northern Italy. In the following, this source of contagion is referred to as \emph{epidemic-between} component, as it concerns the inter-province spread of COVID-19 (see Appendix).

The last portion of cases should be attributed to province-specific conditions which determined the first centres of infections and the initial exposure to the risk of contagion. Such a local component of the overall daily province infections is referred to as \emph{endemic} component, according to terminology introduced in \citet{PaulHeld2011}, and which, in this context, does not implies any epidemiological qualification of the COVID-19 in the population of Italian provinces. We point out that in this paper both terms \emph{epidemic} and \emph{endemic} have been inherited from \citet{PaulHeld2011}, whereas terms \emph{within} and \emph{between} are introduced in order to distinguish between the temporal and spatial terms which \citet{PaulHeld2011} jointly refers to as \emph{epidemic component}.

The statistical model is structured in three parts which contribute to the number of daily infections per province according to the three described components. Each part of the model considers also the presence of heterogeneity both in the temporal and in the geographical mechanisms of contagion, so that the different dynamics existing inside the provinces, both in terms of temporal and geographical mechanisms spread of COVID-19 are modelled.

All analyses presented in this paper were performed using the R software~\citep{Rcite}.

\section{Results}

The estimated model provides useful insights about the evolution and spread of COVID-19 occurrences across the territory of Italy. The most striking evidence is that the epidemic potential across areas is, on average, very strong; however, it is also highly heterogeneous among provinces.

\begin{figure}[p]
\begin{center}
\caption{Maps of the three fitted mean number of infections components as proportions averaged over all days. Colors toward red indicate a relatively higher role of the component; colors toward green indicate a relatively lower role of the component.}
\includegraphics[scale=0.3]{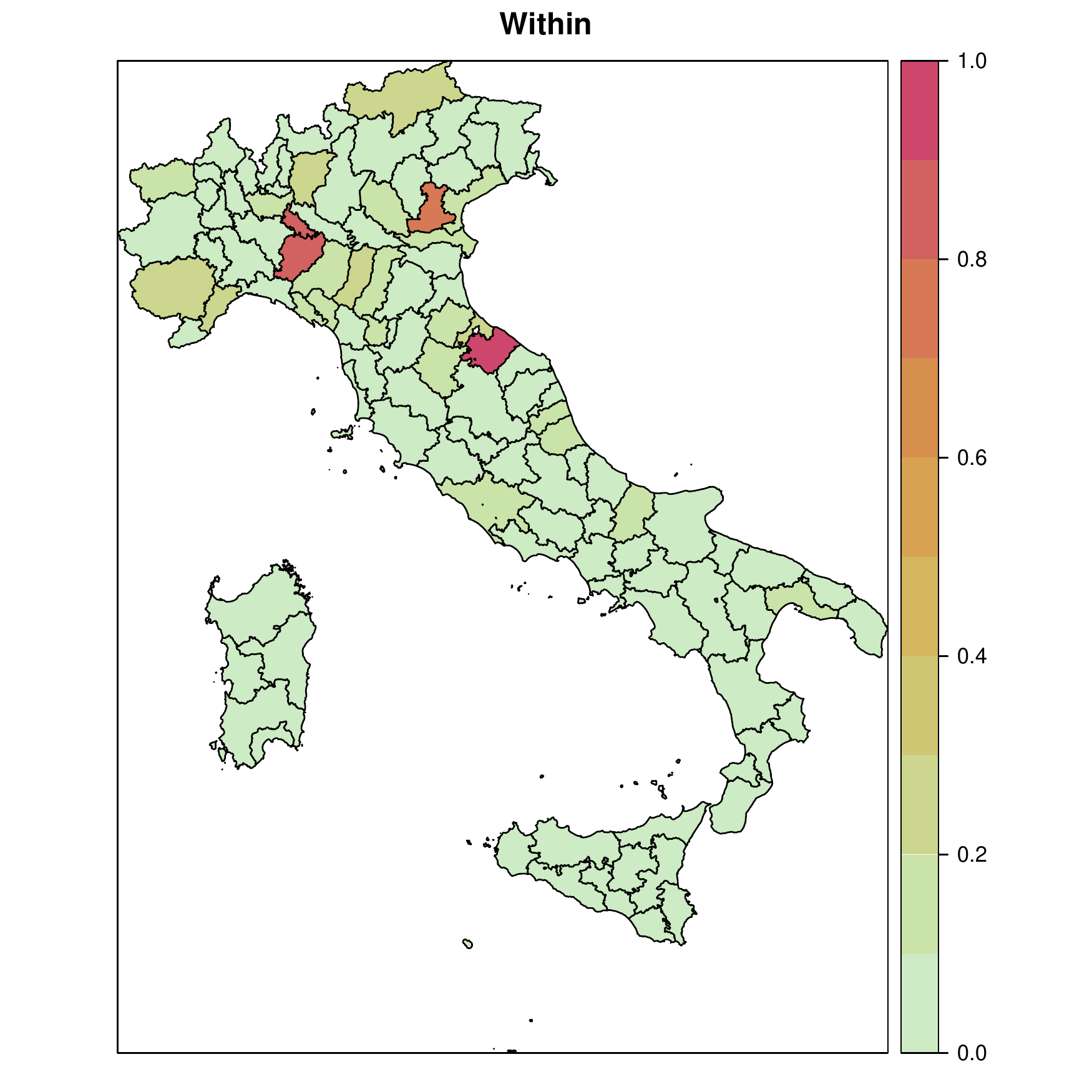}\includegraphics[scale=0.3]{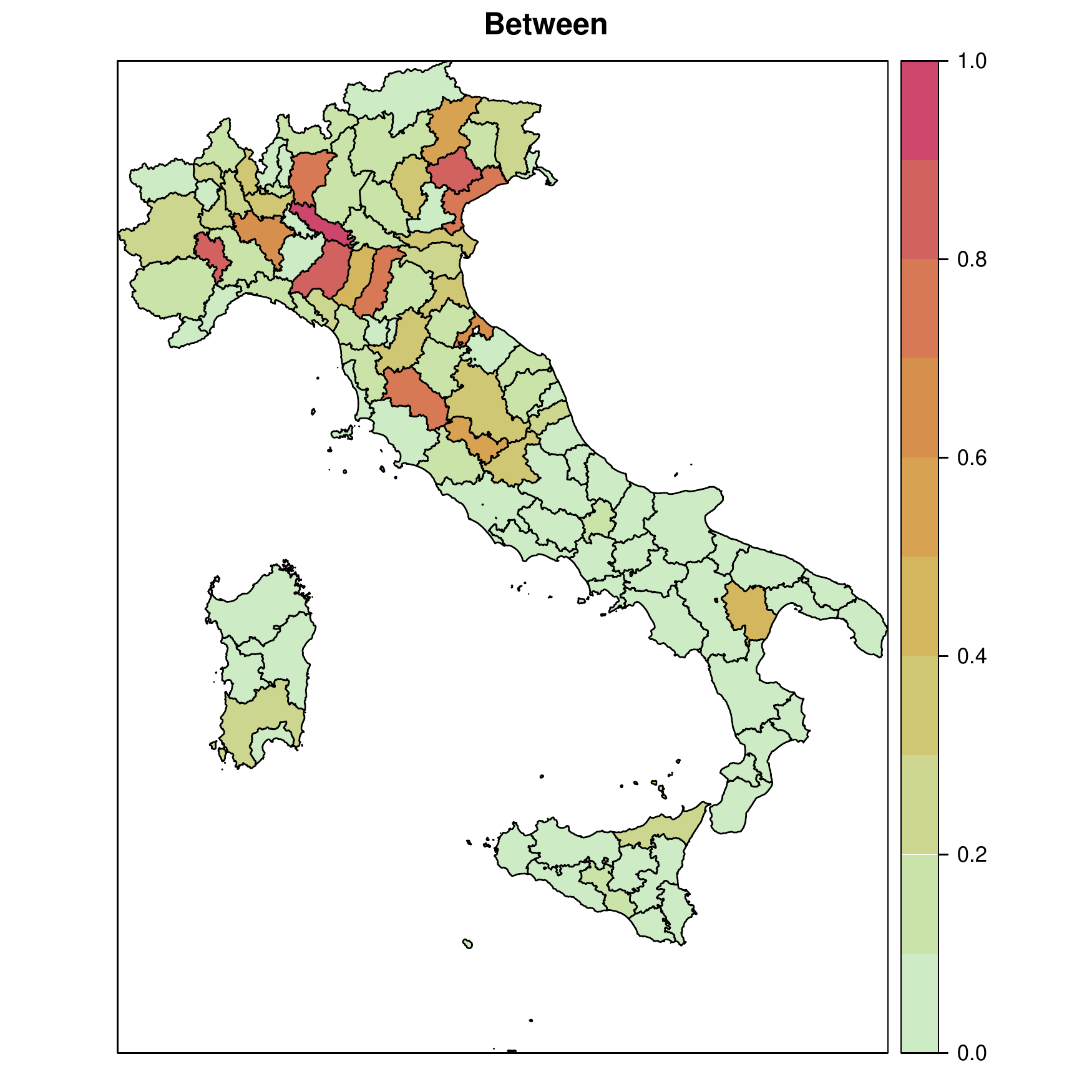}\includegraphics[scale=0.3]{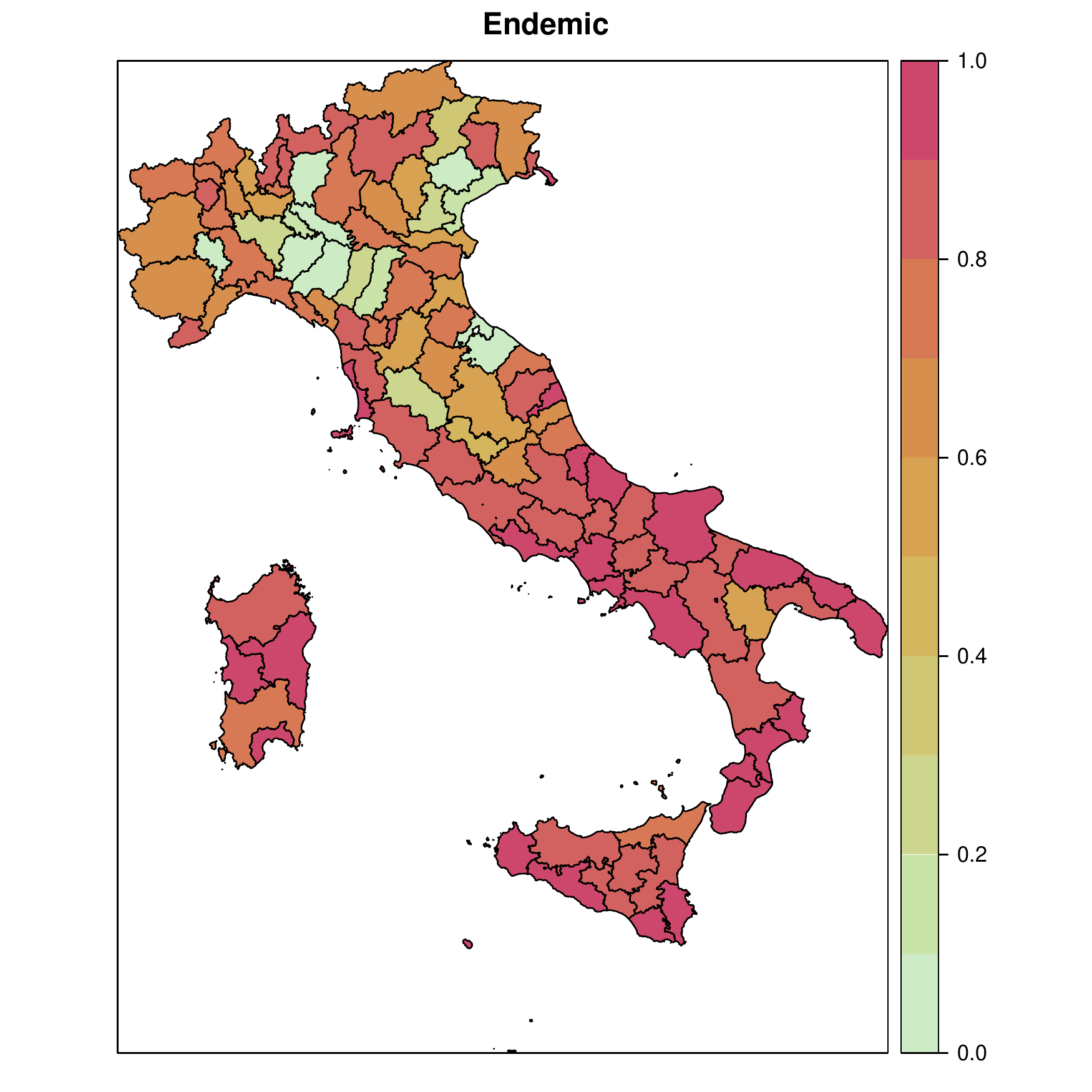}
\label{ComponentsMap}
\end{center}
\end{figure}

To better assess the degree of spatial heterogeneity, Figure~\ref{ComponentsMap} shows three maps depicting the decomposition of the estimated expected number of infections into its three components --- namely the within-epidemic, between-epidemic and endemic contributions --- by provinces. For each province, the three fitted components are expressed as proportions of their sum. We can clearly see that only few provinces, the most affected by the disease until now, are mainly influenced by the local endogenous transmission of the contagion  (map on the left). For a relatively higher number of provinces, mostly located in the north and center parts of the country, a relevant number of cases is instead explained by the transmission from neighboring provinces (map on the centre). For the majority of provinces, located mainly in the south, the contagions follow essentially the endemic trend (map on the right). The comparison among provinces in terms of the three investigated components is robust to spuriousness because the Italian population is entirely and homogeneously susceptible, that is immunologically ``naive'', to this new virus. Moreover, being vaccines still missing, there are no differences among provinces deriving from vaccination coverage.

\begin{figure}[p]
\begin{center}
\caption{Locations of the paradigmatic provinces in the north of Italy.}
\includegraphics[scale=0.6]{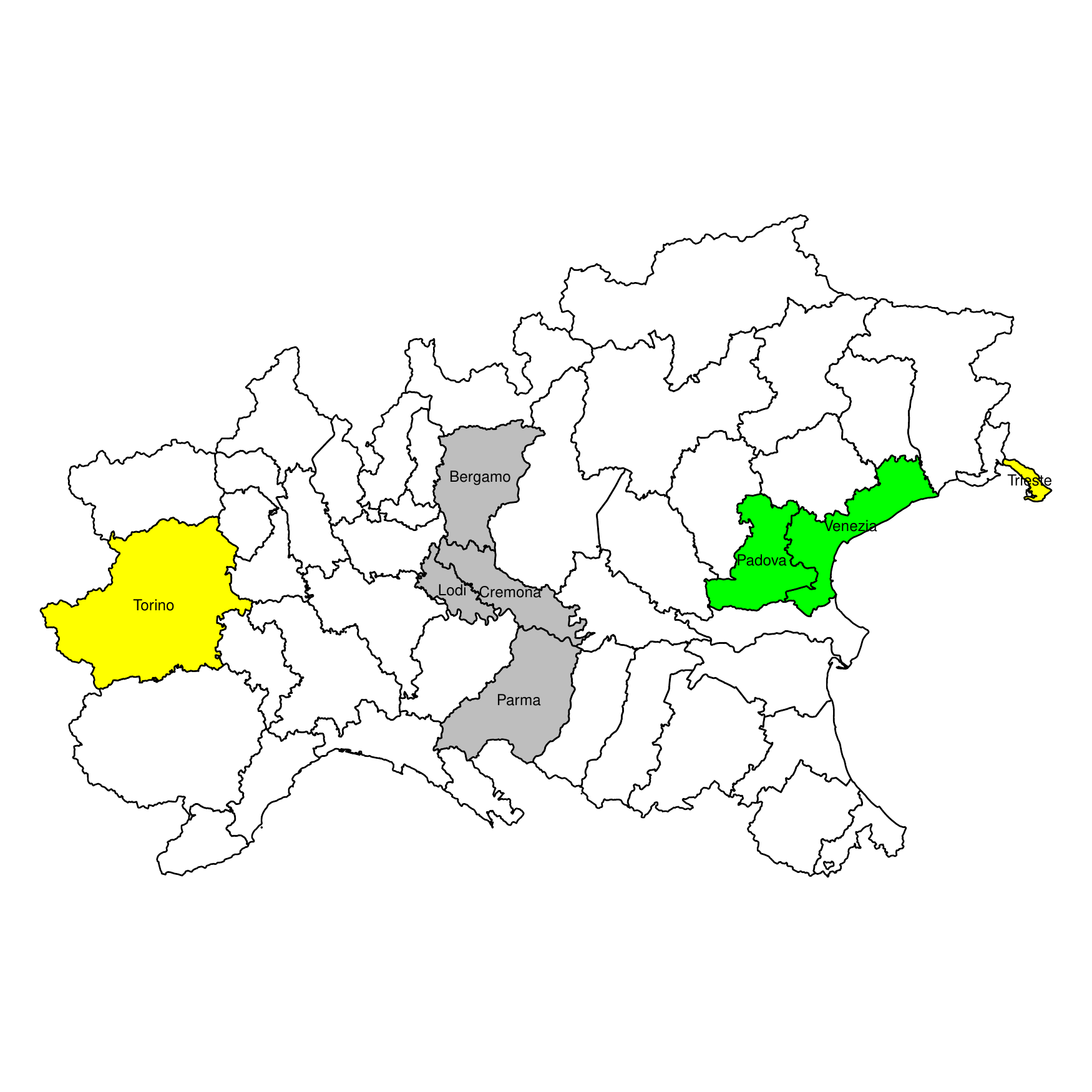}
\label{mappa_paradigm}
\end{center}
\end{figure}

\begin{figure}[t]
\begin{center}
\caption{Fitted means of the three submodels for the eight NUTS-3 regions with a higher number of infections. The dots represent the observed daily counts. The blue area represents the within-epidemic component. The orange area represents the between-epidemic component. The gray area represents the endemic component.}
\includegraphics[scale=0.6]{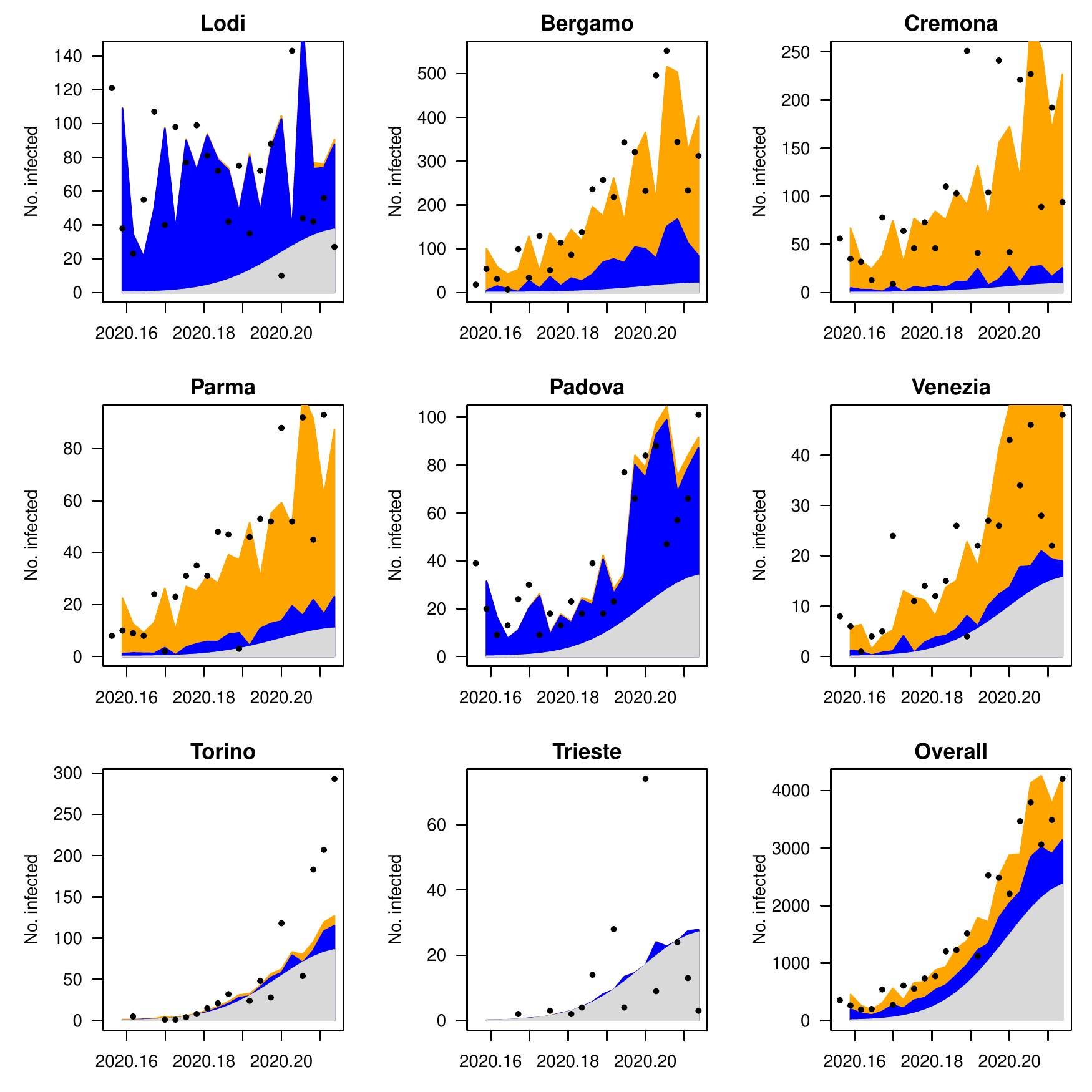}
\label{fittedcomponents}
\end{center}
\end{figure}

To gain further insights about the relative importance of the three components we take into consideration some paradigmatic provinces located in the part of the country with the highest number of cases, that is the northern areas (see Figure~\ref{mappa_paradigm}). In particular, we focus the attention on the northern part of Italy because after the illness started to occur in the country, it took several days to spread throughout the territory, thus leading to an unbalanced spatial distribution of infections. Figure~\ref{fittedcomponents} depicts the mean number of cases estimated by the model along with the observed number of cases for the paradigmatic provinces. First of all, the province of Lodi (Lombardy region), which is the first area that reported cases, sees a predominance of the component related to the internal diffusion of the disease. This evidence agrees with the quarantine to which some of its municipalities underwent since the beginning of the outbreak and with the hypothesis that the epidemic in Italy started right there. The province of Lodi shares the eastern border with the province of Cremona (Lombardy region), that in turn shares the northern border with the province of Bergamo (Lombardy region) as depicted in Figure~\ref{mappa_paradigm}. The latter two are so far some among the most severely affected provinces in Italy. Their proximity with the province of Lodi makes them widely susceptible to the contagion effect between neighboring regions. This suggests that the contagion of infections among inhabitants occurred before containment measures were carried out and that the disease remained undetected before patient 1 was discovered. A similar situation arose for the province of Parma (Emilia-Romagna region), which share the norther border with the province of Cremona. Even here the between-epidemic component is relevant and it may be explained with the strong social and economic ties existing among the areas.
The second in time discovered outbreak of COVID-19 in Italy was detected in a small town in the province of Padova (Veneto region). Even in this case, strong control measures to reduce transmission outside have been carried out, as evidenced by the almost total influence of the within-epidemic component. This shows that there have been no ``return infections'' from the surrounding areas but it does not mean that, before the quarantine, the infection could not have circulated in the neighboring provinces. In fact, the province of Venice (Veneto region), which is geographically, culturally and economically near to that of Padova demonstrates to be affected by these form of interactions.
The cases of the provinces near the national borders are also of particular interest, such as the provinces of Turin (Piedmont region) and Trieste (Friuli Venezia Giulia region), adjoining with France and Slovenia, respectively. These provinces may be affected by the so-called boundary problem for which the analysis of a phenomenon may be biased by the presence of arbitrary administrative geographic borders. Thus, on one side we might lose information regarding what happens beyond the country border, due to different ways of collect information and monitor the diffusion of the disease. One the other side, we might not have a clear overview about cross-border phenomena as job trajectories. In addition, the two considered provinces seem to be not connected with the aforementioned outbreaks originated in Lombardy and Veneto regions. Due to these reasons, the provinces of Turin and Trieste show a limited effect of epidemic components and a prevalence of the endemic component, which is due only to the evolution of the disease over time.
Looking at the national aggregated counts, the exponential evolution of COVID-19 contagions is clear. On the country as a whole, the main component seems to be the temporal trend but also the epidemic components have a not negligible role. The observed decomposition is probably mainly due to the central and southern provinces, which are characterized almost exclusively by the spread over time component.
 
Moreover, to check the ability of the proposed model in explaining the spatio-temporal distribution of the COVID-19 occurrences in Italy, we examine how it is good at predicting the future daily counts of infections. Therefore, we re-estimate the model using the data between 26~February~2020 -- 17~March~2020 and we make prediction for 18~March~2020. The observed total number of cases at 18~March~2020, obtained by aggregating the province data, is 4204; the model provides a prediction of 4191 cases. Therefore, the model underestimates by only 14 units. Naturally, the advantage of the proposed model is that it provides predictions for each individual province (point predictions can be found in Tables~2 and 3 of the  Appendix) but it is clear that the use of predictions at local level can have positive effects on  the prediction of total number at country level. We report (in Figure~2 of the Appendix) the 80$\%$ prediction intervals of the provincial counts of infections at 18~March~2020. Given the brevity of the observed time series, to obtain a reasonable degree of precision we cannot require higher confidence levels. Despite the fact that the intervals are quite wide, the results are promising if compared with the known observed counts. We expect that the level of precision will improve as the data are updated and the observed time series become longer.

\section{Discussion}

In this article, we analyzed, by modelling, the trend of COVID-19 epidemics along time and space. The use of spatio-temporal models can greatly improve the estimation of number of infected and can help the public decision-makers to better plan health policy interventions. 
Italy has viewed a massive spread of the disease with peculiar patterns on the territory. Started from some provinces in the northern area, this serious illness descended down the Boot and is nowadays present in all 107 Italian provinces. Several discrepancies among provinces, especially between north and center-south of the country, persist but coming weeks seem to be decisive to understand the behavior of the contagion trend. Containment measures in Italy have followed an application in three steps: first, some municipalities in Lombardy and Veneto regions underwent to quarantine; second, the entire Lombardy and some provinces in other northern regions (Veneto, Piedmont, Emilia-Romagna and Marche) were isolated from the rest of the country; third, all Italian territories were subjected to a complete lockdown. Such stepwise approach in imposing hard control measures to the entire national territory might have conducted to a temporal shifting of the contagion dynamics.

Emblematic is the case of the province of Lodi, the first area submitted to the quarantine. After an average increment of 85 cases when the illness began to be systematically discovered, in the last 5 days of the time series, which correspond to three weeks after the lockdown, this province seems to have settled the average number of newly discovered contagions equal to 42 cases (so the contagions have been halved). This fact suggests that policies of contagion containment seem to exert a mild success so far in the areas in which there was an effective enforcement of the control measures. Instead, in metropolitan areas where the population density and the more active social behavior make more difficult to comply properly to the recommendations, the whished effects are not yet taking place. This is the case of the province of Milan, which is experiencing an increasing number of cases in last days (from an average of 9 new cases detected of the beginning to an average of 273 new cases of last days). And then there is the case of center and southern provinces, which have viewed a delayed start of the epidemics but also the arrival, in the last two weeks, of flows of people escaped from northern regions undergone to lockdown. As an example, the province of Florence went from an average increment of 3 new cases of the first period to an average increment of 41 new cases of last 5 days, and, similarly, the province of Naples, has gone from an average increment of 4 new cases to an average increment of 29 new cases, confirming the hypothesis of downhill race of the disease along the peninsula. All this has led to the failure in applying homogeneous measures at national level, also due to regional autonomy in force in Italy and to important local differences, which may lead to delays or failures in containing the contagions from a global point of view. This evidence suggests that consider the peculiarities of local territories may be effective in planning specific strategies for enforcing the containment measures established at national level.

These differences in the dynamics of the epidemics in Italy demonstrate the crucial importance of a strong national coordination level for the homogeneous enforcement of control measures, but also reveal how essential predictions at local level are. Since the epidemics started, a frequent question of the public decision-maker is about when the peak of contagions will manifest. Probably a more appropriate request should concern the emergence of different local peaks in different moments, which Italy should expect in next weeks. The dramatic events in northern provinces can serve as a test bench for the health system, offering an overview about what southern provinces might be confronted. Analyses and predictions both in space and time offer a decisive perspective about which areas may be more affected and when, given the time to the decision-makers to intervene on the local policies.

\section*{Contributors}

DG and FS were responsible for statistical modeling, data analysis and writing of the manuscript. MMD and GE were responsible for result presentation, discussion and writing of the manuscript. All authors reviewed, revised and approved the final manuscript.

\section*{Conflict of interest statements}

\begin{itemize}
\item None of the authors of this paper has a financial or personal relationship with other people or organizations that could inappropriately influence or bias the content of the paper.
\item No competing interests are at stake and there is no conflict of interest with other people or organizations that could inappropriately influence or bias the content of the paper.
\end{itemize}

\section*{Acknowledgments}

Authors thank Umberto Agrimi of the \emph{Istituto Superiore della Sanità} (Italian National Institute of Health, Rome), who read a preliminary version of this article and provided several valuable comments and suggestions.

\bibliography{BibPaper}
\bibliographystyle{plainnat}

\clearpage

\appendix{Appendix}

\section{Statistical model}

The data about the spatio-temporal distribution of Coronavirus disease 2019 (COVID-19) infections at province level consist in multivariate count time series whose spatial references are in the form of irregular spatial lattices. Therefore, the proper regression modelling framework for this empirical circumstance is the class of the so-called areal Generalized Linear Models (GLMs). By extending the seminal model originally introduced by \citet{HeldEtAl2005}, \citet{PaulHeld2011} proposed an endemic-epidemic multivariate time-series mixed-effects GLM for areal disease counts, which proved to provide good predictions of infectious diseases \cite[see][among the others]{AdegboyeEtAl2017, Cheng2016}. 

The main equation of the model describes the expected number of infections $\mu_{r,t}$ in a region (province) $r$ at time (day) $t$ as follows:
\begin{equation}
\label{generalModel}
\mu_{r,t}
=\lambda_r\,Y_{r,t-1}
+\phi_r\sum_{r'\neq r}w_{r',r}\,Q_{r',t-1}
+e_r\,\nu_{r,t}\,,
\end{equation}
where $Y_{r,t}$ is the number of infections reported in the region $r$ at time $t$, which follows a negative binomial distribution with region-level overdispersion parameter $\psi_r$. If $\psi_r>0$ the conditional variance of $Y_{r,t-1}$ is $\mu_{r,t} (1+\psi_r\,\mu_{r,t})$, while if $\psi_r=0$ the negative binomial distribution reduces to a Poisson distribution with parameter $\mu_{r,t}$.  

The three terms on the right-hand side of Equation~\eqref{generalModel} correspond to the three components of the model: the \emph{epidemic-within}, the \emph{epidemic-between}, and the \emph{endemic}.

The first component models the contribution of temporal dynamics of contagions to the expected number of infections within region $r$. The term includes the number of infections observed in the previous day (time $t-1$), which affect $\mu_{r,t}$ depending on the value of the coefficient $\lambda_r>0$. As the notation suggests, $\lambda_r$ changes amongst the provinces because of a random effect which allows for heterogeneous behaviour in the dynamics of contagions.

The epidemic-between component models the contagion between neighbouring provinces by including the average incidence of the infections $Q_{r',t-1}$ of provinces $r'$ which are neighbours of province $r$. In particular, the coefficients $w_{r',r}$ in the summation $\sum_{r'\neq r}w_{r',r}\,Q_{r',t-1}$ are positive if either province $r'$ and $r$ share a border or province $r'$ and $r$ share a border with the same province, whereas $w_{r',r}$ is zero otherwise. The coefficient $\phi_r$ determines the magnitude of the effect of inter-province spread of contagion, and changes amongst provinces according to the population as well as to unobserved heterogeneity in the diffusion process.

The last component determine the province-specific contribution to the number of infections, once the temporal and spatial autoregressive effect are accounted for. The term $e_r$ is the population of province $r$, whereas term $\nu_{r,t}$ consists of a national time trend component, and a province-specific effect depending on the share of population over 65, and on a random effect which catches the heterogeneity due to unobserved factors.

\citet{PaulHeld2011} suggested that the endemic and epidemic subcomponents can be modelled themselves through log-linear specifications:
\begin{align} 
 \label{modelSubcomponent2} 
  \log(\lambda_{r,t}) &= \alpha_r^{(\lambda)} + {\boldsymbol{\beta}^{(\lambda)}}^\top \boldsymbol{z}^{(\lambda)}_{r,t} \:, \\
   \label{modelSubcomponent3} 
  \log(\phi_{r,t}) &= \alpha_r^{(\phi)} + {\boldsymbol{\beta}^{(\phi)}}^\top \boldsymbol{z}^{(\phi)}_{r,t} \:,\\
  \label{modelSubcomponen1}
  \log(\nu_{r,t}) &= \alpha_r^{(\nu)} + {\boldsymbol{\beta}^{(\nu)}}^\top \boldsymbol{z}^{(\nu)}_{r,t} \:.
\end{align}

\noindent where the $\alpha_r^{(\cdot)}$ parameters are region-specific intercepts and $\boldsymbol{z}_{r,t}^{(\cdot)}$ represent observed covariates that can affect both the endemic and epidemic occurrences of infections. The varying intercepts allow to control for unobserved heterogeneity in the disease incidence levels across regions due, for example, to under-reporting of actual infections \citep{PaulHeld2011}. Given the regionally decentralized health system in Italy, non-negligible differences in case reporting of COVID-19 infections among Italian regions are very likely, which make the opportunity to have region-specific intercepts very important. Following \citet{PaulHeld2011} region-specific intercepts can be obtained through the inclusion of random effects. In particular, we assume here that
$$
\alpha_r^{(\lambda)} \stackrel{iid}{\sim} N(\alpha^{(\lambda)}, \sigma_\lambda^2), \mbox{ and}  \quad
\alpha_r^{(\phi)} \stackrel{iid}{\sim} N(\alpha^{(\phi)}, \sigma_\phi^2), \quad
\alpha_r^{(\nu)} \stackrel{iid}{\sim} N(\alpha^{(\nu)}, \sigma_\nu^2).
$$
The \citet{PaulHeld2011} model with normally distributed random effects can be estimated through penalized likelihood approaches that have been implemented in the \textsf{R} package \textsf{surveillance} \citep{surveillance}. See \citet{PaulHeld2011} for futher details.

\subsection{Epidemic-within submodel}

Given the brevity of the observed time series, the epidemic-within autoregressive parameter is  assumed to be constant over time and, in absence of useful exogenous covariates, the model of Equation~(\ref{modelSubcomponent2}) takes the form 
$$
 \log(\lambda_r) = \alpha_r^{(\lambda)},
$$
that is, the ``internal'' infectiousness depends only on a spatially varying intercept.

\subsection{Epidemic-between submodel}

Following \citet{meyer2014}, the subcomponent model for the epidemic within autoregressive parameter, Equation~(\ref{modelSubcomponent3}), is here specified as
$$
\log(\phi_{r,t}) = \alpha_r^{(\phi)} + {\beta_3^{(\phi)}}\log{e_r},
$$
which accounts for the fact that the regions may have different propensities to be affected by the other neighbouring regions, and this may depend by their resident population share. The rationale is that the more populated regions tend to be more susceptible to transmission across regions.

\subsection{Endemic submodel}

Since some first recent empirical evidences suggest that the number of COVID-19 infections seems to grow exponentially over time \citep{expon1, expon2}, the endemic component model assessing the temporal dynamic of disease incidence, Equation~(\ref{modelSubcomponen1}), is specified as a second-order polynomial log-linear regression:\footnote{We do not include higher-order terms to avoid the risk of introducing spurious endemic patterns and overfitting.}
$$
 \log(\nu_{r,t}) = \alpha_r^{(\nu)} + \beta_1^{(\nu)}t + \beta_2^{(\nu)}t^2 + \beta_3^{(\nu)}\log(a_r),
$$ 
where $t=1,2,\dots$ is the time in days and $a_r$ is the proportion of inhabitants over 65 years old.

In the global model of Equation~(\ref{generalModel}) the endemic predictor $\nu_{r,t}$ is multiplied by the offset $e_{r}$, which in our case is the regional share of resident population.

\section{Supplementary Results}

\begin{figure}[htb!]
\begin{center}
\caption{Time series plots of daily counts of COVID-19 infections in the Italian provinces between February 26 and March 18.}
\includegraphics[scale=0.60]{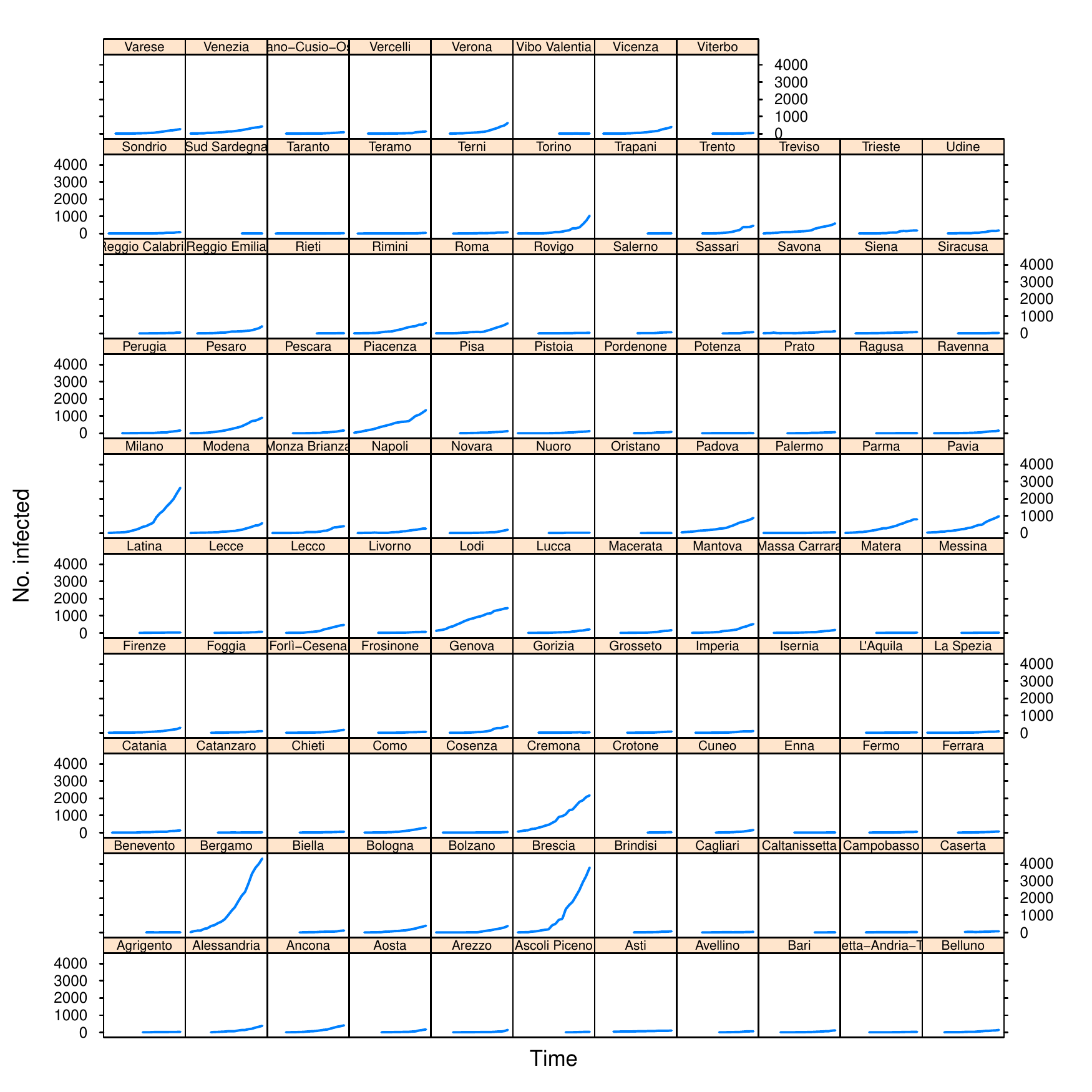}
\label{ProvTS}
\end{center}
\end{figure}

\begin{table}[ht!]
\centering
\caption{Maximum penalized likelihood estimates of model parameters.}
\begin{tabular}{l l r }
\hline
Parameter & Estimate & St. Error$^\circ$ \\
\hline
$\exp(\alpha_r^{(\lambda)})$  & $0.128^{***}$ & 0.029 \\
$\beta_4^{(\phi)}$ & $1.087^{***}$ & 0.392 \\
$\exp(\alpha_r^{(\phi)})$ & $72.551$ & 137.758 \\
$\exp(\beta_1^{(\nu)})$ & $1.791^{***}$ & 0.074 \\
$\exp(\beta_2^{(\nu)})$ & $0.987^{***}$ & 0.001 \\
$\beta_4^{(\nu)}$ & $3.281^{***}$ & 1.195 \\
$\exp(\alpha_r^{(\nu)})$ & $245.462$ & 432.642 \\
$\sigma_\lambda^2$ & $1.387$ & {--} \\
$\sigma_\phi^2$ & $3.419$ & {--} \\
$\sigma_\nu^2$ & $1.310$ & {--} \\
\hline
\multicolumn{3}{l}{\scriptsize{$^{***}p\mbox{-\emph{value}}<0.01$, $^{**}p\mbox{-\emph{value}}<0.05$, $^*p\mbox{-\emph{value}}<0.1$}} \\
\multicolumn{3}{l}{\scriptsize{$^\circ$cannot be estimated for random effects variances $\sigma_.^2$}}
\end{tabular}
\end{table}

\begin{figure}[htb!]
\begin{center}
\caption{Prediction intervals (confidence: $80\%$) of the one-day ahead forecast of the provincial counts of COVID-19 infections (March 18). Each province is identified by the acronym (see also Tables 2 and 3)}
\includegraphics[scale=0.60]{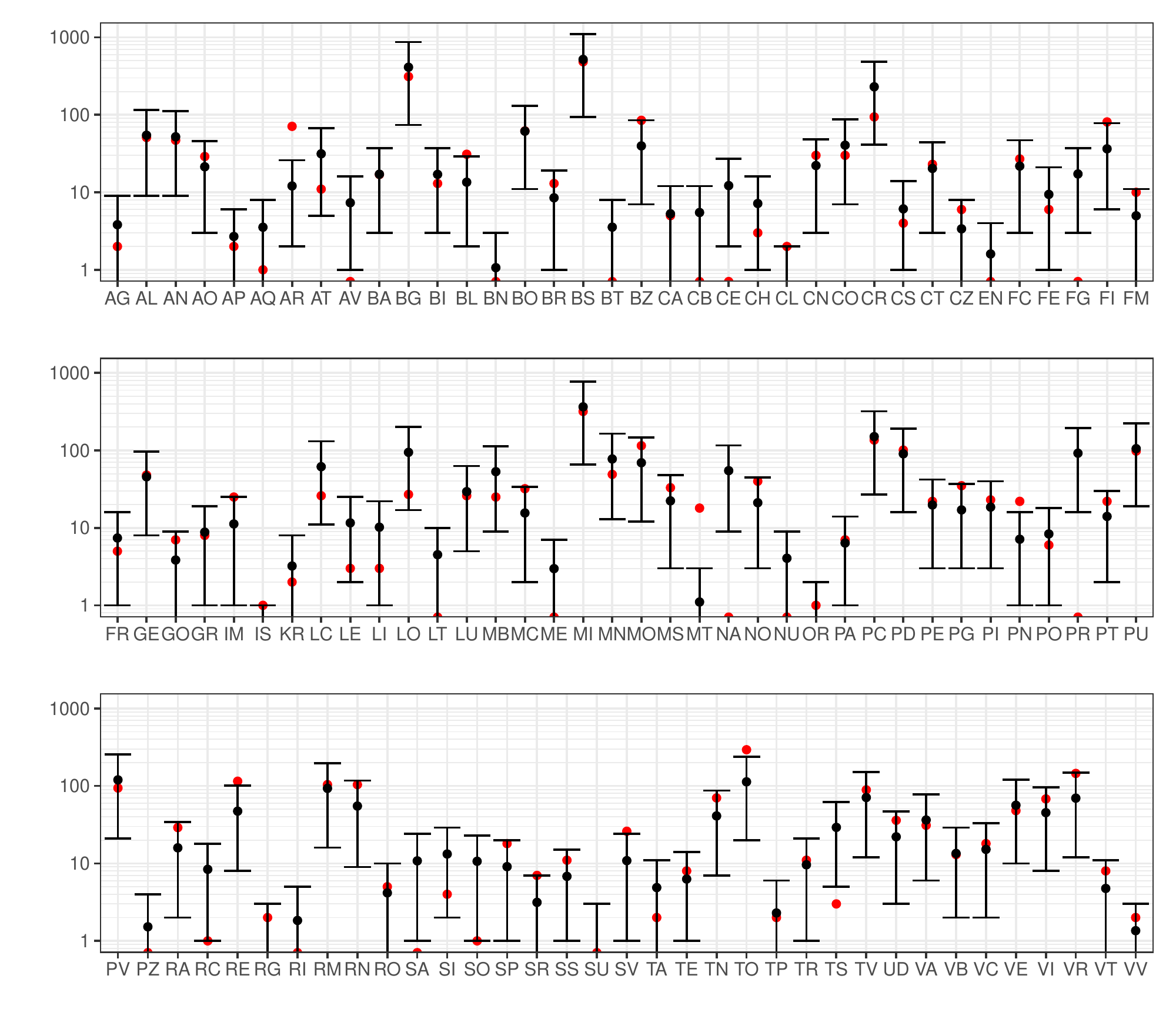}
\label{ProvTS}
\end{center}
\end{figure}

\begin{table}[ht!]
\scriptsize
\centering
\caption{Observed and predicted number of COVID-19 infections at March 18 (continued)}
\begin{tabular}{l l r r }

\hline
Province & Acronym & Observed No. Infections & Predicted No. Infections \\
\hline
Pordenone	&	PN	&	22	&	7.1	\\
Isernia	&	IS	&	1	&	0.3	\\
Biella	&	BI	&	13	&	17	\\
Lecco	&	LC	&	26	&	61.6	\\
Lodi	&	LO	&	27	&	94.4	\\
Rimini	&	RN	&	104	&	55	\\
Prato	&	PO	&	6	&	8.3	\\
Crotone	&	KR	&	2	&	3.2	\\
Vibo Valentia	&	VV	&	2	&	1.4	\\
Verbano-Cusio-Ossola	&	VB	&	13	&	13.4	\\
Monza Brianza	&	MB	&	25	&	53.1	\\
Fermo	&	FM	&	10	&	5	\\
Barletta-Andria-Trani	&	BT	&	0	&	3.6	\\
Torino	&	TO	&	293	&	113	\\
Vercelli	&	VC	&	18	&	15.2	\\
Novara	&	NO	&	40	&	21.1	\\
Cuneo	&	CN	&	30	&	22.2	\\
Asti	&	AT	&	11	&	31.4	\\
Alessandria	&	AL	&	51	&	54.6	\\
Aosta	&	AO	&	29	&	21.4	\\
Imperia	&	IM	&	25	&	11.2	\\
Savona	&	SV	&	26	&	10.9	\\
Genova	&	GE	&	48	&	45.6	\\
La Spezia	&	SP	&	18	&	9.1	\\
Varese	&	VA	&	31	&	36.3	\\
Como	&	CO	&	30	&	40.6	\\
Sondrio	&	SO	&	1	&	10.6	\\
Milano	&	MI	&	318	&	364.8	\\
Bergamo	&	BG	&	312	&	411.8	\\
Brescia	&	BS	&	484	&	519.1	\\
Pavia	&	PV	&	94	&	119.7	\\
Cremona	&	CR	&	94	&	230.2	\\
Mantova	&	MN	&	49	&	77.4	\\
Bolzano	&	BZ	&	85	&	39.7	\\
Trento	&	TN	&	70	&	41	\\
Verona	&	VR	&	145	&	69.5	\\
Vicenza	&	VI	&	68	&	45.2	\\
Belluno	&	BL	&	31	&	13.5	\\
Treviso	&	TV	&	89	&	70.9	\\
Venezia	&	VE	&	48	&	56.3	\\
Padova	&	PD	&	101	&	90.3	\\
Rovigo	&	RO	&	5	&	4.2	\\
Udine	&	UD	&	36	&	22	\\
Gorizia	&	GO	&	7	&	3.8	\\
Trieste	&	TS	&	3	&	29.2	\\
Piacenza	&	PC	&	136	&	150.5	\\
Parma	&	PR	&	0	&	92	\\
Reggio Emilia	&	RE	&	115	&	47.3	\\
Modena	&	MO	&	115	&	69.3	\\
Bologna	&	BO	&	62	&	61.5	\\
Ferrara	&	FE	&	6	&	9.4	\\
Ravenna	&	RA	&	29	&	15.9	\\
Forlì-Cesena	&	FC	&	27	&	21.8	\\
Pesaro	&	PU	&	98	&	105.3 \\	
\hline
\end{tabular}
\end{table}

\begin{table}[ht!]
\scriptsize
\centering
\caption{Observed and predicted number of COVID-19 infections at March 18}
\begin{tabular}{l l r r }

\hline
Province & Acronym & Observed No. Infections & Predicted No. Infections \\
\hline
Ancona	&	AN	&	47	&	52.1	\\
Macerata	&	MC	&	32	&	15.6	\\
Ascoli Piceno	&	AP	&	2	&	2.7	\\
Massa Carrara	&	MS	&	33	&	22.3	\\
Lucca	&	LU	&	26	&	29.4	\\
Pistoia	&	PT	&	22	&	14.1	\\
Firenze	&	FI	&	81	&	36.4	\\
Livorno	&	LI	&	3	&	10.2	\\
Pisa	&	PI	&	23	&	18.5	\\
Arezzo	&	AR	&	71	&	12.1	\\
Siena	&	SI	&	4	&	13.2	\\
Grosseto	&	GR	&	8	&	8.8	\\
Perugia	&	PG	&	35	&	17.1	\\
Terni	&	TR	&	11	&	9.6	\\
Viterbo	&	VT	&	8	&	4.8	\\
Rieti	&	RI	&	0	&	1.8	\\
Roma	&	RM	&	104	&	93	\\
Latina	&	LT	&	0	&	4.5	\\
Frosinone	&	FR	&	5	&	7.4	\\
Caserta	&	CE	&	0	&	12.2	\\
Benevento	&	BN	&	0	&	1.1	\\
Napoli	&	NA	&	0	&	54.8	\\
Avellino	&	AV	&	0	&	7.3	\\
Salerno	&	SA	&	0	&	10.8	\\
L'Aquila	&	AQ	&	1	&	3.5	\\
Teramo	&	TE	&	8	&	6.3	\\
Pescara	&	PE	&	22	&	19.7	\\
Chieti	&	CH	&	3	&	7.2	\\
Campobasso	&	CB	&	0	&	5.5	\\
Foggia	&	FG	&	0	&	17.2	\\
Bari	&	BA	&	17	&	17.1	\\
Taranto	&	TA	&	2	&	4.9	\\
Brindisi	&	BR	&	13	&	8.5	\\
Lecce	&	LE	&	3	&	11.6	\\
Potenza	&	PZ	&	0	&	1.5	\\
Matera	&	MT	&	18	&	1.1	\\
Cosenza	&	CS	&	4	&	6.1	\\
Catanzaro	&	CZ	&	6	&	3.4	\\
Reggio Calabria	&	RC	&	1	&	8.4	\\
Trapani	&	TP	&	2	&	2.3	\\
Palermo	&	PA	&	7	&	6.4	\\
Messina	&	ME	&	0	&	3	\\
Agrigento	&	AG	&	2	&	3.8	\\
Caltanissetta	&	CL	&	2	&	0.8	\\
Enna	&	EN	&	0	&	1.6	\\
Catania	&	CT	&	23	&	20.3	\\
Ragusa	&	RG	&	2	&	1	\\
Siracusa	&	SR	&	7	&	3.1	\\
Sassari	&	SS	&	11	&	6.8	\\
Nuoro	&	NU	&	0	&	4.1	\\
Cagliari	&	CA	&	5	&	5.3	\\
Oristano	&	OR	&	1	&	0.7	\\
Sud Sardegna	&	SU	&	0	&	1 \\	
\hline
\end{tabular}
\end{table}

 \clearpage

\end{document}